\documentstyle[psfig]{l-aa-ps}     % LaTeX A&A  Standard Fonts
\def\ub{$U-B$}
\def\bv{$B-V$}

\def\vi{$V-I$}
\def\br{$B-R$}

\begin{document}

    \thesaurus{11         % A&A Section 11: Galaxies
              (11.13.1;  % Galaxies: Magellanic Clouds
               11.19.4;  % Galaxies: star clusters
               08.08.1;  % Stars: HR diagram
               08.05.1;  % Stars: early-type
               08.02.7;  % Stars: blue stragglers
               08.05.2)  % Stars: emission-line, Be 
             }

\title{On the nature of the blue giants in NGC 330
\thanks{Based on observations taken at the European Southern Observatory,
La Silla, Chile}
}

\author{
Eva K.\ Grebel \inst{1,} \inst{2}
\and
Wm James Roberts \inst{3}
\and
Wolfgang Brandner \inst{4}
}

\offprints{E.K.\ Grebel, Sternwarte Bonn}
\institute{
Sternwarte der Universit\"at Bonn, Auf dem H\"ugel 71,
D--53121 Bonn, Germany, grebel@astro.uni-bonn.de
\and
University of Illinois at Urbana-Champaign, Department of Astronomy, 
1002 West Green Street, Urbana, IL 61801, USA
\and
Center for Astrophysical Science,
Johns Hopkins University,
Box 16329,
Baltimore, MD 21210, USA,
roberts@coyote.pha.jhu.edu
\and
Astronomisches Institut der Universit\"at W\"urzburg, Am Hubland,
D-97074 W\"urzburg, Germany, brandner@astro.uni-wuerzburg.de
}

\date{Received Oct 19, 1995; accepted Nov 15, 1995}

\maketitle

\begin{abstract}
The young SMC cluster NGC 330 contains a number of blue stars that lie
above the main-sequence turnoff found from our isochrone fitting and below 
the position of the blue supergiants. 
We used our own, new spectroscopy and published data on these stars to 
investigate their possible nature.
Problems in interpreting the evolutionary status of the blue giants 
have been found in several preceding studies.
In theoretical H-R diagrams, these stars lie in the rapidly traversed 
post main-sequence gap, similar to the unexpected concentration found by 
Fitzpatrick \& Garmany (1990) in the H-R diagram of the LMC.

We argue that these stars probably are core H burning main-sequence stars
that appear as blue stragglers resulting from
binary evolution as described in the simulations of Pols \& Marinus (1994)
and effects of rapid rotation. 
Many of the blue stragglers are Be stars and likely rapid rotators.
We suggest that there is evidence for the presence of blue stragglers also
in NGC 1818, NGC 2004, and NGC 2100.
We point out that blue stragglers may be a general phenomenon in the CMDs
of young clusters in the Magellanic Clouds and discuss the implications
for IMF and age determinations.  

\end{abstract}

\section{Introduction}\label{sect_intro}

A few years ago, we started a project to determine ages, metallicities, and
reddenings of distinct stellar populations employing a newly developed 
technique of simultaneous multi-colour isochrone fitting (Roberts 1996,
Grebel et al.\ 1994a). When applying this technique to the young cluster
NGC 330 in the Small Magellanic Cloud (SMC), we found a number of stars 
above the main-sequence turnoff, as indicated by our isochrone fit, but below
the location of the blue supergiants (Sect.\ \ref{sect_photdata}). 

When only photometric data are available, these bright
blue stars would usually be considered main-sequence stars, which 
implies that either the cluster is much younger than our 
simultaneous multi-colour isochrone fit shows or forces the conclusion
that there is a considerable age spread in star formation times. 
Spectroscopy is needed in order to
investigate the nature of these stars. We therefore obtained 
spectra to properly classify these stars (Sect.\ \ref{n330data}). 

\begin{figure*}[t]  % 
\centerline{\vbox{
\psfig{figure=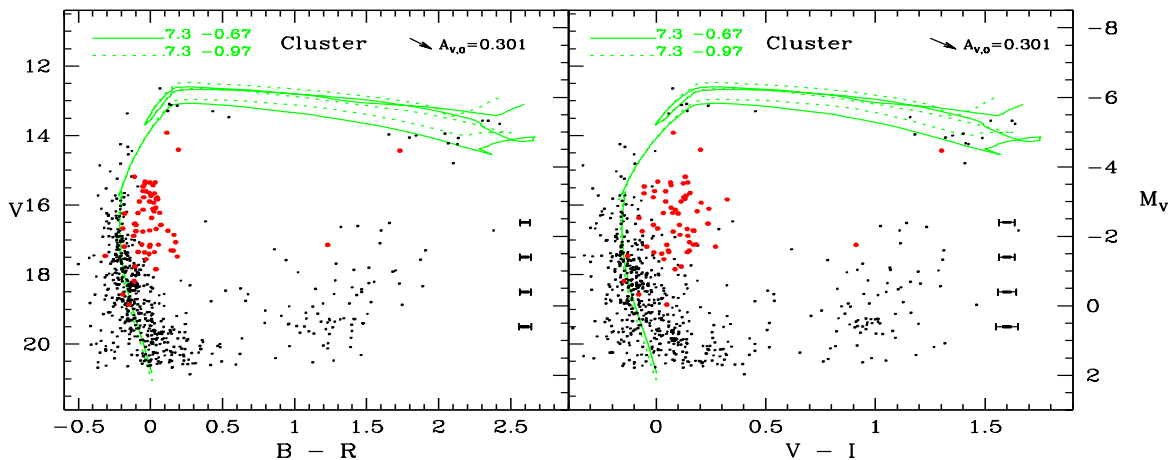,height=6.5cm,width=16cm}
}}
\caption[]{ \label{fn330fjohnccmd}
Colour-magnitude diagrams of NGC 330 for stars within a radius of 22 pc 
around the cluster center. Outside this radius, the number density of stars
is constant; i.e., we no longer expect a significant contribution from 
cluster members. Be stars are indicated by fat dots. Error bars are also
plotted for each diagram.
Superposed are our Geneva isochrones (Schaller et al.\ 1992, Charbonnel et al.\
1993) transformed by us to the
observational plane (Roberts 1996, Grebel et al.\ 1994a). 
These isochrones fit the location of the blue and red
supergiants as well as the main sequence. The main-sequence turnoff is 
located at $V \approx 15.6$ mag. A number of blue stars is visible  
above the main-sequence turnoff defined by the isochrone and below the 
position of the blue supergiants. 
}
\end{figure*}

\begin{table*}[ht]
\caption[Spectral classifications for luminous blue stars in NGC 330]{
\label{tn330tbluefairies} Determinations of spectral types for a number of
blue stars found above the main-sequence turnoff.
The column names have the following meaning: ``Arp'' stands for the 
naming convention introduced for bright stars in NGC 330 by Arp (1959)
on his plates I and II. ``Rob'' denotes the naming convention used by
Robertson (1974). ``FB'' gives the spectral classifications 
summarized in the paper by Feast \& Black (1980), ``CJF'' lists 
the classifications of Carney et al.\ (1985), and ``LMPBC'' comprises 
the classifications carried out by Lennon et al.\ (1995). 
``Above MS''
indicates by ``y'' for yes and ``n'' for no whether stars lie above or 
below the main-sequence (MS) turnoff defined by our best-fitting 
isochrone. ``LDMPM'' refers to radial velocities from Lennon et al.\ (1996).
The rest is self-explanatory. Radial velocities in the 
range from 140 to 155 km s$^{-1}$ indicate likely members of NGC 330. 
}
\footnotesize
\begin{center}
\begin{tabular}{llllllccclll}
\hline  \noalign{\smallskip}
\multicolumn{2}{c}{star ID} & \multicolumn{4}{c}{spectral classification} & & & above & \multicolumn{3}{c}{v$_{\rm rad}$ [km s$^{-1}$]}\\ 
Arp   & Rob & FB & CJF & LMPBC  & this study & V [mag] & B--V [mag] & MS? & \multicolumn{1}{c}{FB} & \multicolumn{1}{c}{CJF} & \multicolumn{1}{c}{LDMPM} \\
\noalign{\smallskip}     \hline  \noalign{\smallskip}
II-13 &A1 &     &      &B0.5 III/Ve &B0-1 III  &$14.81\pm 0.05$&$-0.19\pm 0.10$& y & 145 & & $153 \pm 6$  \\ 
II-14 &A2 &B5 I &B6 I  &B4 Ib       &B5 I      &$13.33\pm 0.05$&$-0.30\pm 0.06$& y & $139\pm 2$ & $147\pm 16$ & \\  
II-31 &A4 &     &      &            &B2-3 IIe  &$14.54\pm 0.02$&$-0.00\pm 0.03$& y & 144 & & \\ 
II-16 &A16&B9 I &A2 I  &            &          &$13.14\pm 0.05$&$+0.06\pm 0.06$& y &  $148 \pm 5$ & $154\pm 8$ & \\ 
II-33 &A25&A1 I &A0 I  &            &          &$13.09\pm 0.02$&$+0.09\pm 0.03$& y & $149\pm 2$ & $155\pm 11$ & \\ 
II-40 &A29&B9 I &      &            &          &$13.12\pm 0.02$&$+0.12\pm 0.02$& y & $139\pm 5$ & & \\
II-35 &A47&B9 I &      &            &          &$12.64\pm 0.02$&$+0.06\pm 0.04$& y & $145\pm 1$ & & \\
I-243 &B2 &     &A7 I  &            &          &$13.80\pm 0.10$&$+0.09\pm 0.11$& y & & $111\pm 17$ & \\ 
I-235 &B4 &     &      &B2 IIIe     &          &$16.21\pm 0.12$&$-0.58\pm 0.13$& n & && $150 \pm 4$ \\
I-234 &B5 &     &      &B2 IIIe     &          &$15.61\pm 0.13$&$-0.03\pm 0.14$& n & && \\ 
I-233 &B6 &     &      &B2 IIIe     &          &$15.59\pm 0.11$&$-0.02\pm 0.12$& n & && \\     
I-232 &B7 &     &      &B2 IIIe     &          &$15.18\pm 0.18$&$-0.75\pm 0.18$& n & && \\ 
I-223 &B11&     &      &A0 II       &          &$15.61\pm 0.09$&$-0.01\pm 0.15$& n & && \\
I-47  &B12&     &      &B2 IIIe     &          &$15.43\pm 0.08$&$-0.05\pm 0.09$& y & 154 & & \\
II-28 &B13&     &      &B2 III/IVe  &          &$15.80\pm 0.03$&$-0.15\pm 0.05$& n & 138 & & \\     
I-222 &B16&     &A0 I  &A0 II       &B9/A0 Ib  &$13.92\pm 0.01$&$+0.08\pm 0.02$ & y & & & \\
I-204 &B18&     &      &O9 III/Ve   &          &$15.74\pm 0.02$&$-0.19\pm 0.03$& n & & & \\
II-46 &B21&     &B4e,n &B2.5 IIe    &          &$14.41\pm 0.02$&$+0.10\pm 0.03$& y & & $113\pm 19$ & \\
II-45 &B22&     &      &B2 IIe      &B3 II     &$14.33\pm 0.02$&$-0.09\pm 0.03$& y & $149\pm 4$ & $149\pm 9$ & \\
I-f   &B24&     &      &B2 IIIe     &B1 IVe    &$15.24\pm 0.05$&$-0.12\pm 0.08$& y & & & \\
      &B28&     &      &B0.2 IIIe   &          &$15.73\pm 0.02$&$-0.19\pm 0.09$& n & & & \\
II-1  &B29&A0 I &A1 I  &            &          &$13.42\pm 0.03$&$+0.05\pm 0.08$ & y & $127\pm 5$ & $122\pm 12$ & \\
II-3  &B30&     &B6 I  &B2 II       &B1 II     &$14.30\pm 0.02$&$-0.14\pm 0.04$& y & $140\pm 4$ & $148\pm 15$& \\
II-4  &B33&     &      &            &B2-3 Ve   &$15.88\pm 0.02$&$-0.16\pm 0.05$& n & & & \\
      &B35&     &      &B2 IIIe     &          &$15.38\pm 0.06$&$-0.09\pm 0.16$ & y & & & \\
II-8  &B37&B5 I &B5 I  &B3 Ib       &          &$13.29\pm 0.05$&$-0.07\pm 0.08$ & y & $148\pm 5$ & $127\pm 26$ & \\ 
II-9  &B38&A1 I &A5 I  &            &          &$13.47\pm 0.10$&$-0.47\pm 0.13$ & y & $151\pm 1$ & $133\pm 14$ & \\ 
I-c   &   &     &      &            &B8 II     &$14.09\pm 0.05$&$+0.18\pm 0.07$& y & & & \\    
I-201 &   &     &A1 I  &            &          &$13.32\pm 0.02$&$+0.14\pm 0.06$ & y & & $103\pm 8$ & \\
I-206 &   &     &      &            &B9 II     &$14.26\pm 0.01$&$+0.03\pm 0.02$ & y & & & \\
I-207 &   &     &      &            &B8 III    &$14.74\pm 0.03$&$+0.12\pm 0.05$& y & & & \\
I-211 &   &     &A0 I  &            &          &$13.81\pm 0.01$&$+0.04\pm 0.05$ & y & & $99\pm 6$& \\
I-218 &   &     &      &            &B1 V      &$15.64\pm 0.23$&$-0.29\pm 0.24$ & n & & & \\
I-219 &   &     &      &            &B9/A0 III &$14.66\pm 0.07$&$-0.09\pm 0.08$ & y & & & \\
I-220 &   &     &      &            &B3 III-IV &$15.16\pm 0.27$&$+0.52\pm 0.35$& y & & & \\
\noalign{\smallskip}     \hline
\end{tabular}
\end{center}
\normalsize
\end{table*}

In addition to our own spectral classifications 
we make use of previously published classifications 
(Sect.\ \ref{sect_specresults}).
These classifications and determinations of surface gravities and effective 
temperatures place these stars where one would not expect to find
so many: in the post main-sequence gap of the Hertzsprung-Russell diagram 
(H-R diagram). 

In the following, we define main-sequence stars as stars in the
core H burning phase. When talking about ``ordinary'' main-sequence
stars we are referring to stars undergoing the usual, undisturbed
main-sequence evolution typical for single stars, to which also the
Geneva evolutionary tracks pertain. According to these models, an
ordinary main-sequence star in the turnoff region of a 20-Myr
isochrone has a $\log g$ value of 3.7 to 3.8.  Of course neither of these
values nor a mere spectral classification alone can tell whether a
star is core H burning or not.

We shall refer to the area redward of the main sequence as
post main-sequence gap or blue Hertzsprung gap. This area lies between
the terminal-age main sequence (TAMS) and the area where the blue
loops of the supergiants occur. Precise size and location of the gap
depend on the stellar models used.  According to stellar evolution
theory, this region is traversed very quickly (see, e.g., Langer \&
Maeder 1995 for time scales) making it highly unlikely to find a
large number of stars there, or even any stars.

In Sect.\ \ref{sectBG_giants} and \ref{sectBG_MS} we discuss whether
the blue stars above the main-sequence turnoff and below the blue supergiant
locus are evolved stars or main-sequence stars. In Sect.\ \ref{sect_exception}
we review indications that NGC 330 is not an exception among young
Magellanic Cloud clusters in containing
blue giants. The implications for the upper IMF and age determinations are
discussed in Sect.\ \ref{sect_imf} and \ref{sect_age}. 

\section{The photometric data}\label{sect_photdata}

We obtained our photometric data 
with the 2.2m MPIA telescope at ESO, La Silla, Chile, on 14 Nov 1992
and on 08 Oct 1993. We used the ESO Faint Object Spectrograph and Camera
(EFOSC II) with a 1024 $\times$ 1024 Thomson chip (ESO \# 19). Pixel
scale was $0\farcs332$ (unbinned) resulting in
field of view of $5\farcm7 \times 5\farcm7$.
We obtained short and long exposures
in Bessel UBVR, Gunn i, and H$\alpha$ with the 2.2m's standard filter set.

We observed several UBV(RI)$_{\rm C}$ standard fields from Landolt (1992)
on 08 Oct 1993 resulting in a total of 47 standard star observations.
The data were reduced using the standard procedures in DAOPHOT II
running under MIDAS (Stetson 1992). We then used  
own programs that correctly remove atmospheric extinction effects (Roberts \&
Grebel 1995) and transform to the UBV(RI)$_{\rm C}$ standard  system 
(Roberts 1996). 

Outside a radius of 22 pc around the cluster center the number of stars 
normalized by area is constant. We therefore consider stars outside this
radius to be mainly field stars. In Fig.\ \ref{fn330fjohnccmd}, only stars
within 22 pc around the cluster are plotted. Our best-fitting isochrone is
indicated by a solid line. For more details on the isochrone fitting procedure
see Roberts (1996) and Grebel et al.\ (1994a). 
Fig.\ 1b in Grebel et al.\ (1994b) shows the
full frame for NGC 330 and the surrounding field. In both diagrams, several 
stars are visible above the main-sequence turnoff and below the supergiant
locus. 
The main-sequence turnoff as given by the isochrone is at $V \approx 15.6$.
Inevitably, the resulting magnitude difference between the loci of the blue
supergiants and the main sequence is $\Delta V \approx -2.5$ mag, which is
expected
from theory and is independent from the set of opacities and the amount of
overshoot used (Stothers \& Chin 1992). 

Note that our isochrone fits the location of the blue and red supergiants.
In past studies, the stars above the main-sequence turnoff defined by our 
isochrone have either been considered ordinary main-sequence stars whose tip
indicates the main-sequence turnoff (Carney et al.\ 1985) or as main-sequence
stars younger than the rest of the cluster, thus indicating an age spread. 
Possible age spreads of 10 Myr for Geneva models (Bomans \& Grebel 1994), 
15 Myr for semiconvective Padua models, and as much as 38 Myr for models 
with full overshoot (Chiosi et al.\ 1995) have been suggested. However,
when fitting the position of the stars above our isochrone turnoff to younger
isochrones there are no supergiants at the positions 
corresponding to the supergiant loci of these isochrones. 
This might be an effect of small-number statistics, though on the other hand
it seems strange that the supposedly ongoing star formation would not have
resulted in the evolution of additional, younger supergiants. 
Obviously, additional data are needed.

\section{The spectroscopic data}\label{n330data}

\subsection{The observations}

Our spectroscopic data were obtained with a Boller \& Chivens
spectrograph attached to the ESO 1.52m telescope on 29 Dec 1993 and on
28 July 1994.  The B\&C spectrograph was equipped with a UV coated
$2048 \times 2048$ Ford chip (ESO \# 24).  The angular pixel scale
was $0\farcs81$/pixel.  We used grating \# 23, covering a
wavelength range from 390 to 780 nm in the first observing run and
370 to 730 nm in the second run. The slit width was 2$''$. With
0.19 nm/pixel, we obtained a resolution of $\approx$ 1000 in the blue.
An order separation filter (WG 360) with a cutoff at 360 nm was used
to prevent second order overlap.

The orientable slit allowed us to observe two target stars 
simultaneously. Because of the wide slit we
often obtained spectra of other stars as well in one exposure. Of course
we did not get all the light
from the other stars since some of them would be positioned on the edge
of the slit. Additional light loss occurred due to guiding errors
(an autoguider was not yet available at the 1.52m telescope) and 
refractory losses at higher airmasses. Therefore no flux calibration
was attempted, though we observed two spectrophotometric standards per
night to determine the response curve of the instrument.
The FWHM of the PSF in
spatial direction varied between $1\farcs 8$ and $2\farcs 8$.
In total we observed 20 stars located in the outer regions
of the cluster (not all of them blue). All spectra had exposure times of
30 min (except for one exposure of 25 min). 

\subsection{Reduction and analysis}

The raw data were reduced in the standard way, i.e., bias subtracted
and flatfielded by a normalized dome flat in order to remove
pixel-to-pixel variations. The individual spectra were extracted,
sky-subtracted, and wavelength-calibrated using the
dispersion relation from the He--Ar comparison spectra.
For the extinction correction the standard La Silla extinction curve
as incorporated into MIDAS was used.
An instrumental response curve was derived in the same way
from the standard stars. Then the target stars were response-corrected
with this curve and rectified.

The poor signal-to-noise ratio in the violet made it difficult
to determine spectral types for the spectra from Dec 93. Therefore
the data from Jul 94 were used whenever possible.
For our spectral classifications we used digital versions of
the OB star atlas of Walborn \& Fitzpatrick (1990) and 
the spectral atlas by Jacoby et al.\ (1984).

\section{Spectroscopic results from our and other studies}\label{sect_specresults}

\subsection{Spectral classifications}

\begin{figure*}  
\centerline{\vbox{
\psfig{figure=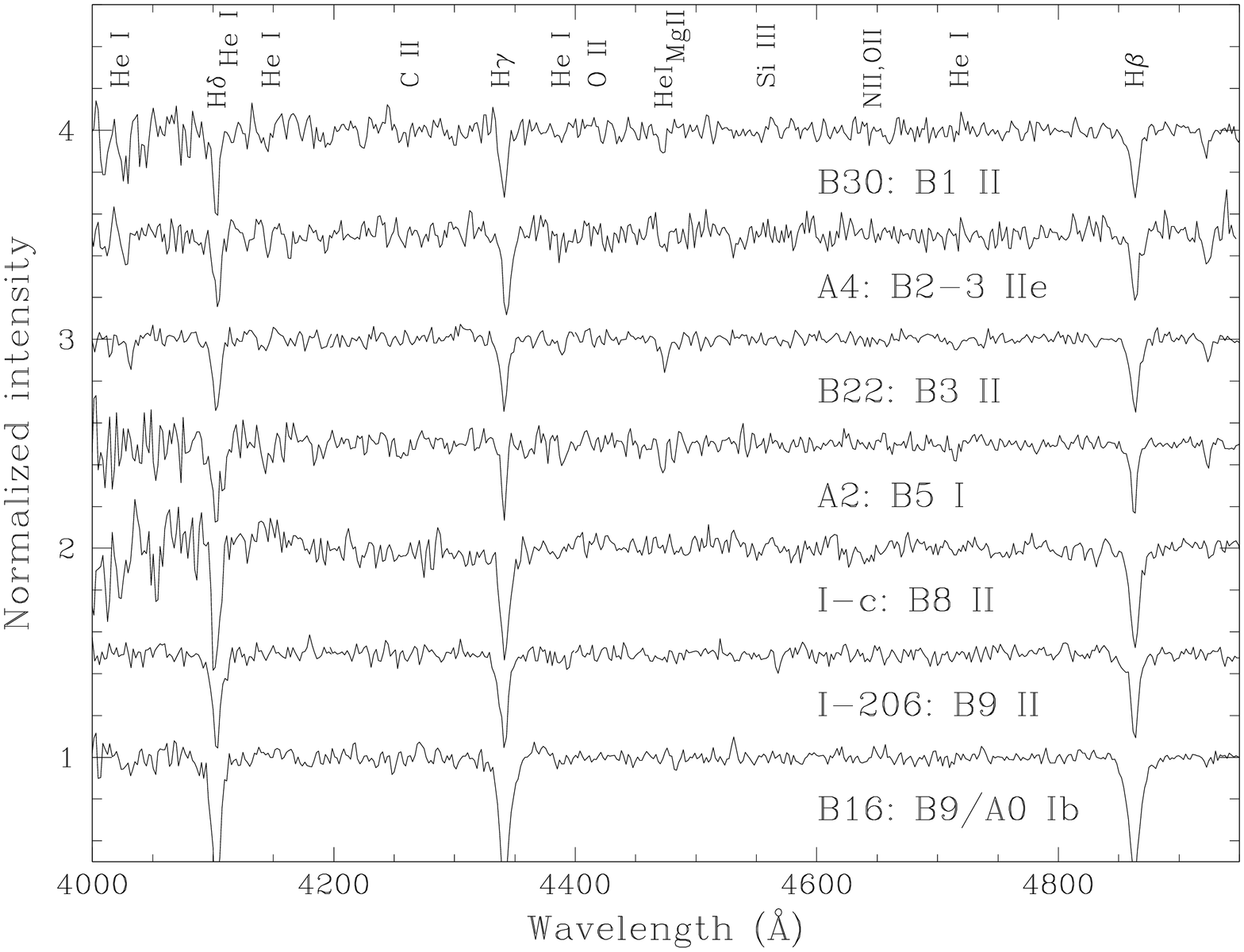,height=11cm,width=16cm,angle=-90}
\psfig{figure=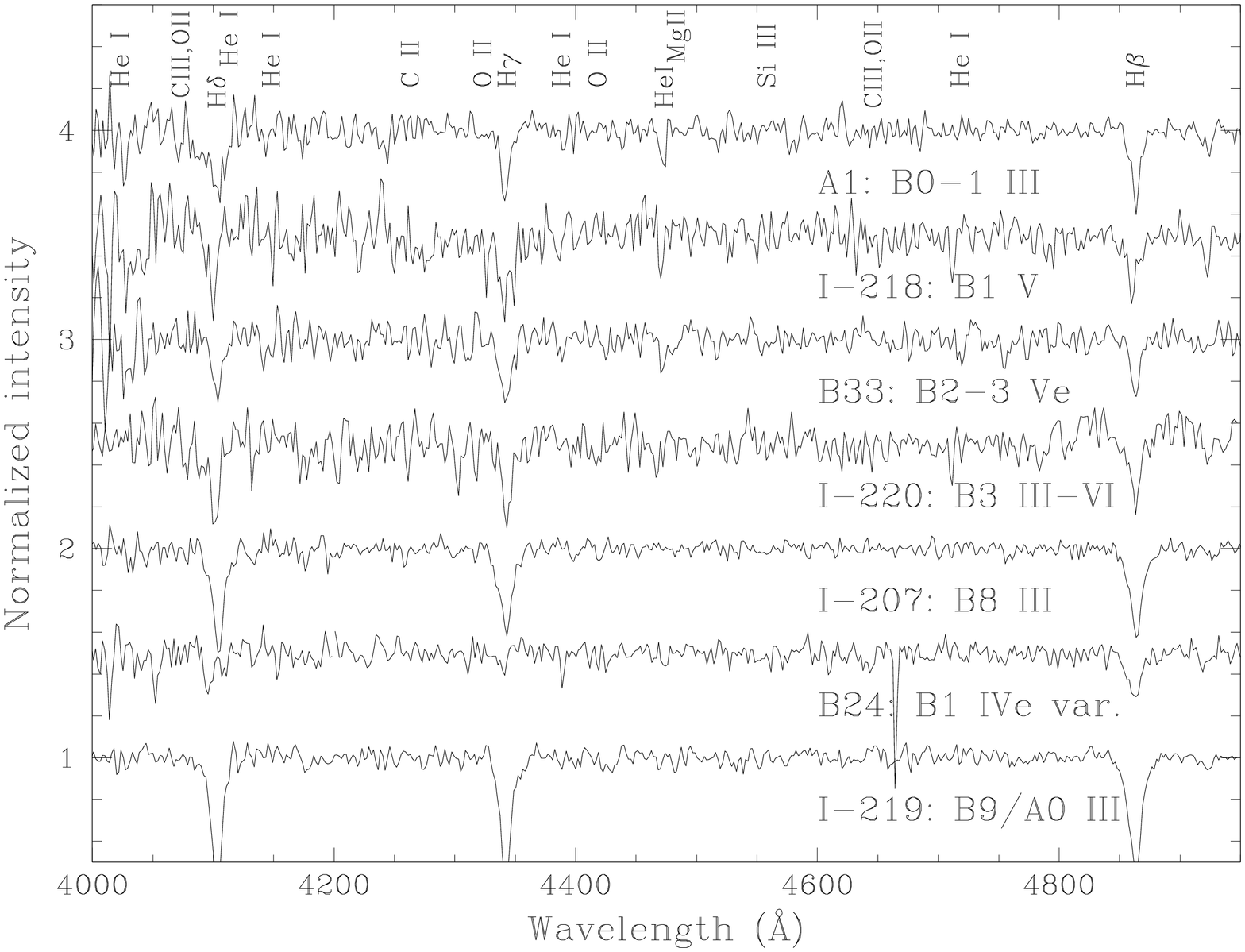,height=11cm,width=16cm,angle=-90}
}}
\caption[]{ \label{n330BFspec1}
The upper panel contains spectra of supergiants and bright giants in NGC 330.
In the lower panel, spectra of giants and main-sequence stars in NGC 330 are
displayed.
}
\end{figure*}

Fig.\ \ref{n330BFspec1} shows our spectra with line identifications
and classifications. Additional spectral
classifications were performed by Feast \& Black (1980), Carney et al.\ (1985),
and Lennon et al.\ (1994, 1995). Combining all these data, we
find that the majority of blue stars with $15.6 \le $ V $\le 14.0$ mag
belong to luminosity class IV, III, and II, while stars brighter than V
$\approx 14$ mag
are supergiants (Tab.\ \ref{tn330tbluefairies}). We shall refer to the
stars of luminosity class IV, III, and II
in the following as {\em blue giants}. The two spectroscopically
classified main-sequence stars 
lie below the main-sequence turnoff
indicated by our isochrones. Thus the main-sequence turnoff determined
from our isochrone fits agrees well with the position of stars identified 
spectroscopically as main-sequence stars, as well as with that of the blue and
red supergiants. 

The supergiants are mostly intermediate- and late-type B supergiants and 
early A-type supergiants. The blue giants span the entire
B-type range. The two B main-sequence stars in our sample are 
early B-type stars (B1 V to B2 V), 
which is not surprising since only spectra of the 
brightest blue stars were taken. 
For a discussion of individual stars, see Sect.\ \ref{sect_indiv}.
The majority of the stars show H$\alpha$ emission.
Spectral types determined by different
authors vary within a few subtypes, which may be due to different 
recording techniques (e.g., resolution, and 
CCD frames versus image tube spectra) and analytical
techniques. Also, spectral classifications are difficult in the metal-poor
SMC due to the fewer/weaker metal 
lines than in luminous blue Galactic stars (Walborn et al.\ 1995, Lennon et al.\
1995). 
The spectra with the highest resolution and best S/N for B stars in NGC 330
are those obtained by Lennon et al.\ (1994).  

\subsection{Determinations of effective temperatures and surface gravities}

Reitermann et al.\ (1990) analyzed high-resolution echelle spectra in the
visual and low-resolution IUE spectra of the B star B30 (naming convention
from Robertson 1974) in NGC 330. In their H-R diagram, B30 lies in the 
post main-sequence gap. Reitermann et al.\ suggest that this star has not yet 
been a red supergiant. Bessell (1991) determined surface gravities 
for B22, B30, and B37 in NGC 330. In both studies, low surface gravities
were found (Tab.\ \ref{tn330surf}).  

Reitermann et al.\ (1990) nicely illustrate in their Fig.\ 2 the uncertainties
in determinations of effective temperatures and surface gravities when 
comparing the results from different methods, namely those based on the 
``reddening-free index'' Q = (\ub$)-0.72\cdot$(\bv), the equivalent width
of Balmer lines, and the S{\sc iii}/S{\sc ii} ionization equilibrium. 
Often $T_{\rm eff}$ values are based on Q alone and very susceptible 
to the effects of individual reddenings and colour excesses. Transformation
equations for determination of $T_{\rm eff}$ as given in Massey et al.\
(1995) furthermore require advance knowledge of the luminosity class. 

However, since the analyses by different authors, which are
in part based on different
data and different methods, agree quite well within the errors, we believe
that the location of the blue giants in the blue Hertzsprung gap is not 
caused by erroneous transformations. 

\begin{table*}[t]
\caption[Surface gravities and effective temperatures for luminous blue 
stars in NGC 330]{\label{tn330surf} Determinations of surface gravities, 
effective temperatures, and masses for
blue stars found above the main-sequence turnoff and below the young
blue supergiants. All stars are identified by the denotations 
introduced by Robertson (1974).
``RBSW'' quotes values from Reitermann et al.\ (1990). The column ``Bessell''
gives his measurements from 1991. ``J\"uttner'' gives J\"uttner's values
(1993), ``CCCW'' stands for Caloi et al.\ (1993), and
``LMPBC''gives the results of Lennon et al.\ (1994), which were replaced
by the revised values of Mazzali et al.\ (1995) when available. 
The errors of the $\log g$ determinations range from 0.1 to 0.5. The errors
of the temperature determinations are between 1000 and 5000 K. Column ``Be''
indicates whether the star was found to show H$\alpha$ emission
(Grebel et al.\ 1992, 1994b, Lennon et al.\ 1994).  
The stars A2 and B37 are supergiants. 
Stars identified as variables by Balona (``Ba'', 1992) or by Sebo \& Wood 
(``SW'', 1994) are listed in the last column. 
}
\begin{center}
\begin{tabular}{cccccccccccc}
\hline  \noalign{\smallskip}
Star ID&\multicolumn{2}{c}{RBSW}&Bessell&\multicolumn{2}{c}{J\"uttner}&\multicolumn{2}{c}{CCCW}&\multicolumn{2}{c}{LMPBC/MLPMC} & & Ba/SW\\
& $\log g$& $T_{\rm eff}$& $\log g$& $\log g$& $T_{\rm eff}$&$\log g$& $T_{\rm eff}$& $\log g$& $T_{\rm eff}$ & Be & variable\\ 
\noalign{\smallskip}     \hline  \noalign{\smallskip}
A01 &      &        &       & & & 3.7   & 28,000  & 4.10  & 28,000 & y & \\
A02 &      &        &       & & & 2.5   & 15,000  & 2.50  & 16,000 & n & \\
B04 &      &        &       & & & 3.9   & 23,000  & 3.80  & 25,000 & y & \\
B05 &      &        &       & & & 3.7   & 19,000  & 3.50  & 22,000 & y & Ba964,$\lambda$Eri\\
B06 &      &        &       & & & 3.7   & 22,000  & 3.40  & 22,000 & y & \\
B07 &      &        &       & & &       &         & 3.80  & 24,000 & y & \\
B11 &      &        &       & & &       &         & 3.25  & 12,000 & n & \\
B12 &      &        &       & & &       &         & 3.50  & 23,000 & y & \\
B13 &      &        &       & & &       &         & 3.60  & 22,000 & y & \\
B18 &      &        &       & & &       &         & 4.50  & 32,000 & y & \\
B21 &      &        &       & & & 3.4   & 22,000  & 2.80  & 20,000 & y & \\
B22 &      &        & 3.2   & 3.22 & 22,700 & 3.2   & 20,000  & 3.25  & 21,000 & y & \\
B24 &      &        &       & & &       &         & 3.80  & 25,000 & y & HV1669\\
B28 &      &        &       & & &       &         & 4.50  & 30,000 & y & \\
B30 & 3.00 & 18,700 & 3.1   &2.91& 18,800& 3.2   & 19,000  & 3.25  & 21,000 & n & \\
B32 &      &        &       && & 3.8   & 28,000  &       &        & ? & Ba347,$\lambda$Eri\\
B35 &      &        &       & & &      &         & 3.50  & 25,000 & y & \\
B37 &      &        & 2.8   & & &2.7   & 17,000  & 2.60  & 18,000 & n & \\
\noalign{\smallskip}     \hline
\end{tabular}
\end{center}
\end{table*}

Caloi et al.\ (1993) analyzed IUE spectra of many of the stars below the
blue supergiant locus 
and used Kurucz models to determine temperatures and surface gravities
(Tab.\ \ref{tn330surf}).
Though they could not determine spectral types, they  
found temperatures and surface gravities of these stars as well as their
position in the post main-sequence gap of the H-R diagram incompatible with
their being main-sequence stars. They pointed out problems in interpreting
the evolutionary status of these stars, a finding that was confirmed by 
Lennon et al.\ (1994). Lennon et al.\ determined temperatures and surface 
gravities from fits of LTE and non-LTE model atmospheres and Robertson's
(1974) BV photometry. 

\subsection{\label{sect_indiv} Results for individual stars}

Caloi et al.\ (1993) find the stars B4, B5, B6, B21, B22, and B30 to lie
on the subgiant branch in their theoretical H-R diagram.  These are statistically 
too many stars for a phase traversed so quickly. They mention that B4 
because of its low temperature resembles an evolved star, which
would mean an age spread of 15 Myr. Masses of these apparently evolved 
stars seem to be
of the order of 10 M$_{\sun}$, which would indicate that mass loss is a 
lot less efficient than usually assumed. 
In J\"uttner's (1993) H-R diagram star B22 lies very close to the TAMS, 
while B30 lies in the post main-sequence gap. J\"uttner
derives a mass of $\approx$ 10 M$_{\sun}$ for B30 and $\approx$ 15 M$_{\sun}$
for B22.

Lennon et al.\ (1994) suggest that some of the supergiants (A2, B37) and bright 
giants (B21, B22, B30) with redder colours may be evolved post red 
supergiants, while A1 and B18 (both bright, but bluer and higher $\log g$) 
could be blue
stragglers. We are here using the notations introduced by Robertson (1974)
when referring to these stars.
Star B18 was re-classified as an O9 giant or main-sequence star (Lennon et al.,
1995). In our H$\alpha$ frames, this star is the brightest
star at the center of a large, elliptical H{\sc ii} region of about $1'$ in 
diameter, part of DEMS 87 (Davies et al.\ 1976), 
located approximately $1\farcm3$ northwest of NGC 330. B18 may be
(one of) the exciting star(s) of the H{\sc ii} region, in which
case it is probably not a member of NGC 330. This suggestion is 
circumstantially supported also by the fact that B18 is outside the
membership radius described in Sect.\ \ref{sect_photdata}.
The comparatively low V magnitude (for a late O star) may be due to higher
extinction in the H{\sc ii} region, which then would amount to a few tenths
of a magnitude. 
It is possible that B18 is more heavily extincted than other stars in
this region because this O star appears a little fainter than the brightest
early B main-sequence stars. 

Lennon et al.\ (1994) find the bright giants or supergiants
B11 and B16 to be very
red and suggest that these stars might be SMC field stars. In our data
the supergiant B16 is quite red as well and, in our photometry,
shows H$\alpha$ emission, while 
B11 does not have red colours and photometrically appears to be a
main-sequence star. 
A number of stars classified as B giants and subgiants by Lennon et al.\
(1995) or by us 
lie in the main-sequence turnoff region, partly even a little below 
(Tab.\ \ref{tn330tbluefairies}).  
Of these stars, B4, B5, B6, B12, B13, and B28 were 
found to show H$\alpha$ emission and are Be stars (Grebel et al.\ 1992,
1994b, Lennon et al.\ 1994, Grebel 1995, Mazzali et al.\ 1995). 
We classify star B33 as a main-sequence Be star below the main-sequence
turnoff. The more distant (and probable non-member) star I-218 is a
main-sequence star without Balmer emission.
Judging from their $\log g$ values, B28 and B18 may also qualify as a 
main-sequence stars.  Many stars above the turnoff are Be stars 
(e.g., A1, A4, A15, A22, A39, B7, B12, B17, B21, B22, B24, B35, 
Tab.\ \ref{tn330tbluefairies}, Grebel 1995), some of which were identified
as $\lambda$ Eri variables or eclipsing binaries (Balona 1992, Sebo \&
Wood 1994, Grebel 1995). 

Chiosi et al.\ (1995) plot the stars of Caloi et al.\ (1993) in an 
H-R diagram together
with main-sequence and blue-loop boundaries for different types of 
evolutionary models. They find that when using models with full overshoot,
which considerably widens the main sequence and shifts the TAMS to the red,
all stars are located within the permissible boundaries of the TAMS 
except for B21, B22, and B30, which still lie in the post main-sequence
gap.

There are a number of  
stars above the main-sequence turnoff that do not show Balmer emission,
or did not show Balmer emission at the times they were observed.   
Some of these stars tend to have as blue colours as the main-sequence
stars beneath them, while the Be stars 
generally stand out through their red excess in the redder
colour indices, as one would expect. 

We also note a decline in number density of stars above the
main-sequence turnoff as determined spectroscopically and by isochrone
fitting. This decline can be seen clearly in the \br\ and \vi\ CMDs
(Fig.\ \ref{fn330fjohnccmd}).

We shall now discuss the possible evolutionary status and nature of
those unexpected blue stars above the main sequence and below the area
of the blue supergiants in NGC 330.  There are two possibilities: 
(1) evolved giants or (2) core H burning main-sequence stars.

\section{Blue giants = evolved stars?}\label{sectBG_giants}

We have to consider that the blue giants in NGC 330
may be evolved stars. As mentioned already, 
to find any stars in the subgiant region of the H-R diagram 
at all is unlikely because evolution
here proceeds rapidly (e.g., Langer \& Maeder 1995), and to find so many
is statistically nearly impossible according to current evolutionary models. 
It can be excluded that all of these stars are subgiants. 

\subsection{Evolved field stars}

Of course NGC 330 is heavily contaminated by the surrounding young
field population. Our blue giants might simply be
older, evolved field stars that we see superimposed on the cluster.  
However, the number-density distribution of blue giants is concentrated
toward the cluster, and several of the blue giants are radial velocity
members of NGC 330. We estimate that stars in the velocity range of 140
to 155 km s$^{-1}$ belong to NGC 330 (Tab.\ \ref{tn330tbluefairies})
following the measurements of 
Feast \& Black (1980), Carney et al.\ (1985), and Lennon et al.\ (1996), 
which make A1, A4,
A16, A29, A47, B4, B12, B22, and B30 radial velocity members.
Potential additional members with radial velocities at the lower end of
the distribution are A2, A29, and B13. 
We therefore exclude 
that the blue giants are merely evolved field stars seen in superposition. 

Radial velocity measurements for
B37 and B38 by Feast \& Black (1980) support their being members of NGC
330, while Carney et al.'s (1985) results would make them field stars.
The stars B2, B21, B29, I-201, 
and I-211, all of which are supergiants except for B21,
are probably not members of NGC 330 (which is further supported
by their distances from the cluster). 

\subsection{Post-red supergiants}

While the positions of the supergiants A2 and B37 in the H-R diagram 
(Caloi et al.\ 1993, Lennon et al.\ 1994) agree with Lennon et al.'s 
suggestion that they are He burning, evolved post-red supergiants,
the blue loops of the Geneva models 
do not reach far enough to the blue for the bright giants B21, B22, and B30.
Missing opacity could 
account for missing blue loops for less massive stars (Stothers \& Chin 1994).

The atmospheric composition of post-red 
supergiants
should have changed due to admixture of CNO-cycle products (Walborn
1988), so high-resolution spectroscopic analyses should be able to
clarify their evolutionary status.  
The Ledoux criterion plus semiconvection work well for low-metallicity 
stars (Stothers \& Chin 1994), though adoption of
the Ledoux criterion has little effect on stars in their main-sequence phase
(Stothers \& Chin 1992).
However, for metal-poor red supergiants the Ledoux criterion will lead to 
normal C/N ratios because of shallow convection envelopes. The same holds
for the blue supergiant phase. Thus alteration of surface composition
is not necessarily to be expected.

Fitzpatrick (1991) found a large
spread in N equivalent widths among B-type supergiants in the LMC, and
Fitzpatrick \& Bohannan (1993) suggest that 10 to 20\% of the stars in
their sample are pre-red supergiants.  
For some Galactic blue supergiants 
it has been suggested that they suffered partial mixing of
CN-cycled gas near the main sequence (Gies \& Lambert 1992, Lennon
1994, Venn 1995), i.e., these stars, too, show CNO
abundances different from unevolved B main-sequence stars.

High-resolution spectroscopy of blue supergiants in the Magellanic
Clouds generally showed significant He and N enrichment, while C and O
were depleted (Kudritzki et al.\ 1987, Lennon et al.\ 1991, Humphreys
et al.\ 1991, Fitzpatrick \& Bohannan 1993, Walborn et al.\ 1995),
which has been attributed to these stars being post-red supergiants.
This indicates that even in metal-poor environments such as in the
Magellanic Clouds the surface composition changes in general when 
evolving through the supergiant phase.

As Caloi et al.\ (1993) point out, high-resolution
studies of B30 in NGC 330 do not show He or N enrichment (Reitermann et al.\
1990). The same was found for B22 (J\"uttner et al. 1993,
Tab.\ \ref{tCNO}). It therefore does not
seem very likely that these blue giants are post-red supergiants.

In a high-resolution study of the narrow-line Be stars
A1 and B4, Lennon et al.\
(1996) find these stars highly enriched in nitrogen, while no helium
enrichment and no carbon deficiency is observed as would be expected 
from supergiant evolution (Tab.\ \ref{tCNO}). 

\subsection{Schwarzschild convection \& Case-B evolution?}

When the Schwarzschild criterion for convection is used, a main-sequence
star can evolve after core H exhaustion to become a blue supergiant before
having passed through the red-supergiant phase (Case-B evolution, 
Chiosi \& Summa 1970 and discussion in Fitzpatrick \& Garmany 1990). 
A star of 15 M$_{\sun}$ would be a Case-B blue supergiant at
an age of about 11 Myr following the evolutionary calculations for Case-B
by Brunish \& Truran (1982). 

A comparison of Brunish \& Truran's Fig.\ 3 and the H-R diagram given by Caloi 
et al.\ (1993) and Lennon et al.\ (1994) shows that the location of the 
``blue giants'' is compatible with the position of blue supergiants in
Brunish \& Truran's models. However, if we consider the blue giants to
have evolved in the sense of Brunish \& Truran, we face the problem of
how to explain the presence and location of the ``normal'' supergiants, both
blue and red. Besides, we found NGC 330 to be about twice as old as the
above-quoted model from fits to Geneva isochrones, which of course 
are based on different model parameters such as a certain amount of 
convective core overshooting leading to higher ages, a different
metallicity, and current opacity models. If we were to assume an age
spread to explain the presence of such massive Case-B B supergiants, we
would again be confronted with the problem of explaining why within one 
cluster two different types of stellar evolution and 
convection could take place. 
It does not seem likely that the Case-B scenario is at work in NGC 330.

\subsection{Dependence of derived $\log g$ and $T_{\rm eff}$ on rotation}

The effects of rapid rotation may change the inferred spectral 
classification (Slettebak et al.\ 1980). As Collins (1987) points out,
deviations from spherical symmetry caused by rapid rotation imply that
there is no longer a global $T_{\rm eff}$ and $\log g$. Instead, one can only
determine a specific $T_{\rm eff}$ and a specific $\log g$ that 
is valid for the observed inclination angle. Both quantities vary across
the stellar surface depending on co-latitude. From pole-on inclinations
to equator-on, the spectroscopic gravity may differ by a factor of ten,
and the photometrically derived effective temperature may vary by several
thousand degrees (see Collins's (1987) Fig.\ 2, see also Slettebak et al.\
1980). This should be kept in mind
when looking at the $T_{\rm eff}$ and $\log g$ values listed in
Tab.\ \ref{tn330surf}. If rapid rotation plays a r\^ole here, then these
values do not necessarily imply that these stars are evolved. 
The effects of rotation will be discussed in detail in the next section.

\section{Blue giants = main-sequence stars?}\label{sectBG_MS}

\subsection{Ordinary main-sequence stars}

Luminous 
core H burning main-sequence stars may spectroscopically appear as giants,
as has long been known (e.g., Massey et al.\ 1995 and references therein).  
However, ordinary core H burning stars would not be found in the blue
Hertzsprung gap of the H-R diagram. 
Also the sudden decline in number density above the main-sequence turnoff
indicates that the blue giants may be different from main-sequence stars.

For the blue giants to be ordinary main-sequence stars, the main sequence
needs to be a lot wider than given by the models of the Geneva group -- 
so wide that it covers that part of the blue
Hertzsprung gap in which the blue giants are found. Following the 
suggestion by Tuchman \& Wheeler (1990), Chiosi et al.\
(1995) demonstrate that this can partially be achieved by 
full overshoot models, for which several but not all
of the blue giants in NGC 330 lie 
close to the TAMS, while the blue Hertzsprung gap and the blue supergiant loop
are shifted further to the red.  
Full overshoot is, though, an extreme hypothesis that does not fit well with
modern mixing length theories (e.g., Grossman et al.\ 1993). 
Both the Padua group and the Geneva group have meanwhile 
adopted a moderate amount of overshoot that seems to fit observations
in the Milky Way and in the Magellanic Clouds best. Thus,
our blue giants do not seem to be ordinary main-sequence stars in the sense
previously defined. 

Bessell \& Wood (1993) pointed out that the location of the blue giants
at lower effective temperatures could also be reproduced when using the 
evolutionary tracks of Doom (1985). These models use the Roxburgh criterion
for convection and overshoot, are more centrally concentrated, and reach 
higher luminosities and lower effective temperatures. The location of the
TAMS is in agreement with the location of the blue giants. Spin-up, proposed
by Bessell \& Wood as a possible cause for the Be phenomenon due to 
rotationally induced mass loss, is fostered by these models. These models
might be a valid explanation for what we observe. However, if with anticipation
of the discussion in later sections we take into
account that each cluster has a certain fraction of binary stars, in the 
models of Doom these stars would have to appear shifted even further 
to the red,
and as we shall see later, that does not agree with what is observed in
Galactic open clusters with binaries (s.\ Pols \& Marinus 1994).  

\subsection{Rapid rotators}

Models taking into account high rotational velocities may also be able to
account for (part of) the blue giants. Many of the blue giants are Be stars,
thus supposedly rapid rotators. We will discuss the implications of the
presence of Be stars below. 

Rotational effects were 
studied by Collins \& Smith (1985, s.\ also Maeder \& Peytremann 1970, 1972) 
and Collins et al.\ (1991)), who show
that the amount of main-sequence widening depends, apart from rotational
velocities, on whether stars rotate as rigid bodies or differentially.  
The stronger the differential rotation, the wider the main sequence. The
main sequence of differentially-rotating stars broadens particularly
for stars of later types (A,F), while 
the effects are less pronounced in the steep part described by 
B-type stars. 

As Collins et al.\ (1991) have shown, the observed brightness of a B star
is (among other things) a function of its rotational velocity and its
inclination to the line of sight. A pole-on rotator (inclination angle 
$i < 30\degr$) is displaced upwards from the main sequence of ``non''-rotating 
stars and will be much (up to a magnitude) brighter than other
B stars of the same spectral type but at different inclination 
angles. The larger the inclination angle (maximum $90\degr,$ equator-on;
common definition for equator-on: $i > 60\degr$, Collins 1987)
of a B star toward us,
the redder the star becomes, and, though still brighter than a non-rotating
star, less luminous. Therefore, a star with a given, high rotational speed 
describes what Collins et al.\ (1991) call the ``rotational displacement fan''
in a colour-magnitude diagram (s.\ their Fig.\ 1), its actual colour 
and brightness depending on its inclination angle.

As Collins et al.\ explain, rotational distortion causes a pole-on 
rotator to present a larger surface area including the high-temperature
polar regions free of limb darkening and gravity darkening. Therefore 
such a star appears brighter than a non-rotating star seen pole-on. 
Rapidly rotating stars seen equator-on exhibit very strong limb darkening
in their polar regions and, through rotational distortion, have more
extended, cooler equatorial regions than non-rotating stars. Both effects
make them fainter than non-rotating counterparts. The latter also causes
them to appear redder. 

Rapid rotation also influences the interior of the rotating star, in
essence causing a lower surface temperature, and stars may appear even
more shifted to redder colours than in models considering only 
rotational distortion-induced changes in the atmosphere. 

The positions of the Be stars in NGC 330, which mostly are redward of the
main sequence in our CMDs (Figs.\ \ref{fn330fjohnccmd}),
may to a large extent be caused by rotational displacement as just explained.
Stothers \& Chin (1992) 
suggest that either strong convective envelope overshoot
or fast interior rotation may account for the H-R diagram of NGC 330 and
point out the high fraction of presumably rapidly rotating Be stars in NGC 330.
For most of the stars without Balmer emission and vertically above the main 
sequence their height above the main sequence is in accordance with what one
might expect for pole-on rotators. More about that in the next section.

\subsection{Be stars?}

Most of the blue giants have been identified as Be stars (Grebel et al.\
1992, 1994b, Lennon et al.\ 1994, Grebel 1995, Tab.\ \ref{tn330surf}
and \ref{tn330tbluefairies}). As already mentioned,  
not all of the blue giants currently show H$\alpha$
emission, but the Be phenomenon is known to be variable and episodic. 

Most (Galactic) Be stars are rapid rotators (Slettebak 1988). 
The effects of rotational distortion were just explained in the 
preceding section. 
Be stars in addition often show an infrared excess due to free-free
emission in their disks, which already becomes visible in red filters
of the visual wavelength range (Fig.\ \ref{fn330fjohnccmd}). 

The position of the majority of the blue giants in the
H-R diagram can be explained by their being Be stars and thus shifted to lower
effective temperatures and lower surface gravities due to effects of
rotation and (infra-)red excess. 

Since NGC 330 is unusually rich in Be stars (Grebel et al.\ 1992, 1994b),  
we will now discuss the hypothesis that all blue giants
in NGC 330 are Be stars, even though some do not show Balmer emission at the
moment. If all blue giants were Be stars, this would imply that the stars
vertically above the main sequence must be pole-on rotators. 
We assume that the rotational axes of these stars would have to be within about
$30\degr$ of the line of sight so that limb darkening and the small
solid angle of the equatorial regions would make the stars
significantly brighter and bluer (and smaller angles would be
significantly more effective).  The solid angle within $30\degr$ of
{\em both} poles is about $2\cdot2\pi\cdot(1- cos(30\degr)) =
4\pi\cdot0.134$, 
so the probability of a Be star's
having such an orientation is roughly 1 in 8. In our CMDs of NGC 330,
we see nine blue giants that lie more or less vertically above the 
main sequence. If these are pole-on rotators, we expect approximately
63 non-pole-on Be stars. 
Rapid rotation may affect both B and Be stars. Furthermore, it has to be 
considered that pole-on rotators will not only be present above the
main sequence, but also along the main sequence just as the Be stars
span a range of magnitudes. Also, for our comparison of pole-on and 
non-pole-on rotators to be meaningful we must confine ourselves to
stars only within
the magnitude range that may give rise to the observed nine vertically
displaced blue giants. Thus we need to count B stars and Be stars
within one magnitude below the main-sequence turnoff (and the Be stars
above it) to establish a valid base of comparison. The number ratio
is approximately 60 supposedly non-pole-on B and Be stars versus nine possible
pole-on rotators above the main sequence. (At this point, we note again
that these nine stars above the main sequence do not show H$\alpha$ 
emission.) Considering the uncertainties of the numbers, 
incompleteness, and crowding, we cannot rule out that those
nine blue giants are pole-on rotators.  

Statistically, 
the only possibility for all blue giants in NGC 330 with or without
current H$\alpha$ emission to be Be stars is that {\em all} B stars
in NGC 330 would have to undergo the Be star phenomenon at some point of
their evolution.  If all bright B stars 
are rapid rotators this would suffice to explain the presence of the 
vertically displaced blue giants. 
As rapid rotators and/or Be stars,
the blue giants may well still be core H burning. 

It is not yet known if the B-type stars in NGC 330 are rapid rotators.
Lennon et al.\ (1994) and 
Mazzali et al.\ (1995) determined rotational velocities
for a large number of Be stars in NGC 330. However, these measurements
are based on the H$\alpha$ emission line, which originates in the 
circumstellar disk around the Be stars and may not be representative
for the rotational velocity of these stars. Reitermann (1989) determined
$v \sin i$ values for seven B type stars in NGC 330. The $v \sin i$ values
range from 45 km s$^{-1}$ to $> 100$ km s$^{-1}$. Since the inclination
angles are not known, we do not know what the
rotational velocities are. At the current time, it cannot yet be 
verified whether or not B-type stars in NGC 330 are rapid rotators.
Based on fitting their models, Stothers \& Chin (1992) suggest that stars 
in NGC 330 suffer fast interior rotation. Balona (1992) estimates from
the mean periods and magnitudes of the 19 $\lambda$ Eri variables that
he found in NGC 330 that the mean equatorial rotational velocity of these
stars may be 375 km s$^{-1}$, a relatively high value as compared to Galactic
$\lambda$ Eri variables. Fast rotation may facilitate the formation of Be
stars and short-periodic Be variables such as $\lambda$ Eri stars.  

We point out in this context  
that NGC 1818 and NGC 2004, two young clusters in the LMC that also 
seem to contain blue giants, have comparatively many fewer Be stars
(Grebel et al.\ 1994b, Sect.\ \ref{sect_exception}).

\subsection{Blue stragglers and binary stars}\label{sect_BS2}

Those blue giants vertically
above the main sequence could also be blue stragglers. By definition,
blue stragglers are stars that appear to have ``straggled'' in their
main-sequence evolution and, while still being core H burning 
main-sequence stars, lie
above the main-sequence turnoff given by the age of the cluster in question
(Stryker 1993).  Blue stragglers have been found in all kinds of 
stellar aggregates, from old globular clusters down to young OB 
associations (Mermilliod 1992).

Blue stragglers may be the result of delayed formation (star formation
over an extended period of time) or of delayed evolution (binaries
with mass transfer, Eggen \& Iben 1988). In the first case, the blue
stragglers are likely to be single stars. 
Should the blue giants found vertically above the main sequence be the result
of delayed formation, we would expect them to show spectral characteristics
typical for earlier types of stars. In the case of NGC 330, where the
earliest stars on the main sequence found from isochrone-fitting
are of type B0 to B1, we would 
consequently expect the most massive younger stars to be O stars. This
however is not what we observe. 

Thus the delayed-evolution scenario appears to be more likely. In fact
there is mounting evidence that blue stragglers result from binary
interactions (Stryker 1993). 

Collier \& Jenkins (1984) present model calculations for Case-B mass transfer,
i.e., mass transfer through Roche-lobe overflow while the primary 
component is already a red giant. To quote, the ``two most notable features
resulting from the inclusion of the binary population are the splitting of the
main sequence, and the presence of a sequence of blue stars near the main
sequence but {\em above} the cluster turn-off.'' (Collier \& Jenkins 1984). 
This effect is visible in our CMDs and even more pronounced in NGC 1818
(Will et al.\ 1995). 

As Collier \& Jenkins point out, for stars
with primaries more massive than 2.5 M$_{\sun}$, Case-A mass transfer is
more prevalent than Case B. In Case-A evolution, the system is so close that
Roche-lobe overflow occurs already on the main sequence (Kippenhahn \& 
Weigert 1967). Following Kippenhahn's \& Weigert's model calculations, the
age of NGC 330 is compatible with its containing very massive secondaries (mass $>$
11 M$_{\sun}$) that
would appear as blue stragglers. The binary system in this case
looks like a semi-detached
Algol-type system, where the less massive component fills its Roche lobe
and appears as overluminous subgiant. 
On the other hand, de Loore \& Vanbeveren (1994) argue that with decreasing 
metallicity, the importance of Case-A mass transfer decreases as well. 

Tuchman \& Wheeler (1990) present a scenario in which the Hertzsprung gap can
be filled if a main-sequence star gets enriched in He by a more massive
red-supergiant companion through Roche-lobe overflow. The main-sequence
star becomes a peculiar blue giant while the more massive companion may
become a supernova or a neutron star. 

As mentioned before, though,  high-resolution studies of two bright giants
in NGC 330 (Reitermann et al.\ 1990,
J\"uttner et al.\ 1993) did not find evidence for He enrichment, nor
for N enrichment (compare with J\"uttner et al.'s results for field stars,
Tab.\ \ref{tCNO}). That would seem to exclude the possibility that we are
seeing the effects of homogeneous evolution of massive stars leading to the
formation of OBN blue stragglers (Maeder 1987, Beech \& Mihalas 1989). 
These two stars with their very low surface gravities may, however,
be coalesced binaries or a similar result of massive close binary evolution. 
On the other hand, Lennon et al.'s (1996) finding of 
significant N 
enrichment in A1 and B4 without the accompanying C and O 
depletion (compare to the results of Rolleston et al.\ 1993, who used 
the same method; Tab.\ \ref{tCNO}). The authors suggest
binary mass transfer or rotationally induced mixing for these two 
stars with low v sin i values. They might be pole-on rotators,
but a number of discrepancies are yet to be solved. 
For discussions, see Luck \& Lambert (1992) and Lennon et al.\ (1996).

\begin{table*}[t]
\caption[CNO Abundances for main sequence and evolved stars]{\label{tCNO} 
Comparison of CNO abundances in the Milky Way, the LMC,
the SMC, and NGC 330. 
``RD'' denotes the results of Russell \& Dopita (1992), ``LL''
stands for Luck \& Lambert (1992), ``BSSM'' for Barbuy et al.\ (1991), 
``JT'' for Jasniewicz \& Th\'evenin (1994), ``MBP'' for Meliani et al.\
(1995), ``GA'' for Grevesse \& Anders (1989), Dufour for Dufour (1984),
and ``RD90'' for Russell \& Dopita (1990).
``DBFL'' denotes Dufton et al.\ (1990), ``JSWB'' indicates J\"uttner et al.\
1993, ``RDFHI'' stands for Rolleston et al.\ (1993), ``LDMPN'' for Lennon 
et al.\
(1996). For the B stars, the third row of the headers gives names of individual
stars, and the fourth row gives their spectral classifications. Note that
with the exception of NGC 330, no nitrogen lines were found in the SMC B 
main-sequence stars. The carbon abundances for all stars may be affected
by non-LTE effects. Typical errors for Magellanic Clouds abundances are 
$\pm0.2$ or $\pm0.3$ for individual stars. When names of individual stars
are not listed, the abundances refer to mean values.
Since spectroscopic abundances are still subject to many uncertainties, 
only results obtained with the same models and methods are directly 
comparable.
}
\footnotesize
\begin{center}
\begin{tabular}{cccccccccccc}
\hline  \noalign{\smallskip}
\multicolumn{9}{c}{A,F,G, and K supergiants} & Sun & \multicolumn{2}{c}{H{\sc ii} regions} \\
\noalign{\smallskip}     \hline  \noalign{\smallskip}
Element & Milky Way & \multicolumn{2}{c}{LMC field} & \multicolumn{2}{c}{SMC field} & \multicolumn{3}{c}{NGC 330} & & LMC & SMC \\ 
$\log \epsilon$(M) & LL   & RD   & LL   & RD   & LL   & BSSM & JT & MBP & GA & 
\multicolumn{2}{c}{Dufour/RD90} \\
                   &45 stars&4 stars&15 stars&5 stars&11 stars& 3 stars & 10 stars & 5 stars & & & \\
\noalign{\smallskip}     \hline  \noalign{\smallskip}
C                  & 8.15 & 8.04 & 7.97 & 7.73 & 7.65 & 7.7 & 8.0 & 7.7 & 8.6 & 7.90 & 7.16 \\
N                  & 8.47 &      & 8.36 &      & 7.72 & 6.9 &     &     & 8.0 & 7.07 & 6.55 \\
O                  & 8.61 &      & 8.68 &      & 8.21 & 7.9 &     &     & 8.9 & 8.13 & 8.37 \\
\noalign{\smallskip}     \hline  \noalign{\smallskip}
\multicolumn{12}{c}{B giants and main-sequence stars} \\
\noalign{\smallskip}     \hline  \noalign{\smallskip}
& Milky Way & NGC2004 & \multicolumn{3}{c}{SMC field} & 
\multicolumn{2}{c}{NGC 346} & \multicolumn{4}{c}{NGC 330} \\ 
 & DBFL & JSWB & \multicolumn{2}{c}{RDFHI} & JSWB & 
\multicolumn{2}{c}{RDFHI} & \multicolumn{2}{c}{JSWB} & 
\multicolumn{2}{c}{LDMPN} \\
 &      & D12 & AV304 & IDK-D2 & AV175 & 11  & 637 & B30 & B22 & A01 & B04 \\
 &      &     & B0 V  & B2 III & B III & B0 V & B0 V & B1 II & B3IIe& B0.5III/Ve & B2IIIe \\
\noalign{\smallskip}     \hline  \noalign{\smallskip}
C & 8.2 & 8.00 & 6.9  &   7.1  & 7.68 &$<6.9$& 6.8 & 7.78& 7.79& 6.9 & 6.9 \\ 
N & 7.8 & 7.54 &$<6.9$& $<7.1$ & 7.40 &$<7.6$&$<7.2$&6.92& 6.97& 7.7 & 7.7 \\
O & 8.9 & 8.14 & 8.2  &        & 8.26 &  8.0 & 8.0 & 7.79& 7.98& 8.0 & 8.0 \\
\noalign{\smallskip}     \hline
\end{tabular}
\end{center}
\normalsize
\end{table*}

Pols (1994) presents detailed results for Case-A and Case-B mass transfer
between massive close binaries. It is likely that these interactions result
in one of the components becoming a (comparatively low-mass) He star, which
then would appear blueward of the ordinary core H burning main sequence
(blue interloper). 
While we see some fainter stars displaced to the blue of the hydrogen
main sequence, these stars may simply be there because of large photometric
errors and scatter for fainter magnitudes. 
Spectroscopy of these stars would be useful.

Studying the short-periodic variability of stars in NGC 330, Balona
(1992) found three early-type eclipsing binaries, one of them probably
a contact system. 
In a study of long-periodic variability of stars in the surroundings of
NGC 330, Sebo \& Wood (1994) found one eclipsing binary in the region of
the blue giants (HV 1669 or B24, a B1 IVe star, s.\ Tab.\
\ref{tn330surf}) and three other binaries,
one of them a contact W UMa-type star. Of these, HV11348 was identified as 
Be star (Grebel 1995, suspected already by Carney et al.\ 1985). Both 
stars also lie above the main sequence. Thus, there is observational 
proof for the presence of at least two binaries among the blue giants and 
Be stars. 

Pols \& Marinus (1994) performed Monte-Carlo simulations of binary-star
evolution in young open clusters and compared their results to Galactic
open clusters. They consider four different types of mass transfer and 
assume a binary fraction dependent on the initial mass-ratio distribution
(overall 25 to 30 \% binaries among B-type stars).
For young open clusters, their models reproduce the observations
very well. The binaries cause the main sequence to be widened, or even
a second, binary main sequence to appear. 
Blue stragglers form as result of mass
exchange in close binaries, which may also lead to the formation of Be stars. 
The blue stragglers can either appear as single
stars with fully coalesced components, or have stripped main-sequence
companions, He-star companions, or white dwarfs or neutron stars, depending
on the closeness of the system and the initial masses and mass ratios of 
the components. The blue stragglers are rapid rotators. Stars
rejuvenated by accretion of matter often appear not only brighter but also
bluer than the main sequence.

The simulations of Pols \& Marinus elegantly explain the features we see
in the CMDs and H-R diagrams 
of NGC 330. Combining the previously described effects
of rapid rotation and multiplicity with various forms of mass exchange,
they account for a widened main sequence, for blue giants
that emerge as blue stragglers, for blue giants vertically
above or even to the blue of the main sequence, and for (some of the)
Be stars among the blue stragglers and fainter main-sequence stars. 

Additional observational tests can be carried out to check predictions of
the models. Apart from the spectroscopic 
survey for He stars suggested above, X-ray data of the NGC 330 
region should be investigated for evidence of B or Be stars with
neutron-star 
or white-dwarf companions, which show a different spectral signature
than X-ray flares originating in the disks of Be stars (Cassinelli \&
Cohen 1994). The models
of Pols \& Marinus also predict that for many binaries radial-velocity 
variations are so small that they may escape detection, and in fact 
{\em no} indications for binarity were found in previous radial velocity studies
(Feast \& Black 1980, Carney et al.\ 1985). Long-term photometric monitoring
of NGC 330 similar to the four-year study of the surrounding field population
carried out by Sebo \& Wood would give further clues about the nature of 
various types of binaries in NGC 330 in addition to the short-term variability
studied by Balona (1992).  

\section{Is NGC 330 an exception?}\label{sect_exception}

NGC 330 is exceptional in at least one respect: it contains the highest
fraction of Be stars of any young cluster investigated so far (Grebel et 
al.\ 1994b). 

However, NGC 330 is not unique
in containing what we have named blue giants. The young
LMC cluster NGC 2100 seems to contain several stars that could qualify
as ``blue giants''. B\"ohm-Vitense et al.\ (1985)
analyzed IUE spectra for a number of blue stars in NGC 2100.
The stars C13 and B20 (Robertson's numbers, Robertson 1974) 
have quite low surface gravities though B\"ohm-Vitense 
et al.\ refer to B20 as a main-sequence star. From their placement in the
H-R diagram the stars B20, C13, and C31 could be ``blue giants'' since all of
them lie in the post main-sequence region that is traversed very quickly.
Admittedly, these are very few stars in comparison to NGC 330. Additional
candidates should be investigated. 

Analyzing IUE spectra of blue stars in the young LMC cluster
NGC 2004, Caloi \& Cassatella (1995) find about eight
stars lying in the post main-sequence gap. 
In a CMD, these stars lie at the apparent upper end of the main sequence,
just as in NGC 330. Indeed, if we fit an isochrone to CMDs of NGC 2004
(Fig.\ 1 in Grebel et al.\ 1994b) we find the main-sequence turnoff at lower
magnitudes in good agreement with where Caloi \& Cassatella find the 
first probable main-sequence stars. 

Our own photometry of the young LMC cluster NGC 1818 (Grebel et al.\ 1994b,
Grebel 1995) as well as the study 
by Will et al.\ (1995)
indicates that here as well there is a number of ``blue giants'' above
the main-sequence turnoff and below the position of the blue supergiants.
In fact, the CMDs of Will et al.\ show a forked ``upper main sequence''
with one group of stars shifted toward red colours, while the other
group of stars looks like a slightly blueward extension of the main sequence.

The H-R diagrams
for the young field population of SMC (Garmany \& Fitzpatrick
1989) and LMC (Fitzpatrick \& Garmany 1990) show a possibly related
effect: The post main-sequence gap is not, as one would expect from stellar
evolution theory, empty but instead well filled with stars. It is quite 
possible that this is at least partly
the same phenomenon we observe in NGC 330 and several
other young Magellanic Cloud clusters. We emphasize that in selecting their
stars, Fitzpatrick \& Garmany (1990) excluded all known emission-line 
objects, while in NGC 330 the majority of stars 
filling the Hertzsprung gap shows Balmer
emission. Possible sources for erroneous placement in the Hertzsprung gap
of the Magellanic Cloud field stars are discussed by Tuchman \& Wheeler
(1990). 

It also has to be emphasized that all currently widely available evolutionary
tracks and isochrones consider slow-rotating, single star evolution only.

In Galactic open clusters, widened main sequences, binary main sequences, and
blue stragglers are well-known phenomena and common to clusters of
all ages (Mermilliod 1992). It is hardly surprising that the same
effect should exist in the Magellanic Clouds.  
We suggest that blue giants or blue stragglers may be a common feature in 
young Magellanic Cloud clusters (and possibly in clusters of all ages). 
This possibility should be carefully investigated since it has significant
implications for the determination of ages and IMFs in the Magellanic Clouds.

\section{Re-evaluation of the upper slope of the IMF}\label{sect_imf}

The true nature of the stars at the apparent upper main sequence is
of great significance for the determination of IMFs. If the 
blue giants are all rapidly rotating single stars and single Be stars,
the upper IMF will include stars of erroneously high masses since the 
inclination angle of a rapid rotator may shift it by up to one magnitude
in M$_V$, which in turn implies a higher luminosity in the H-R diagram and a 
higher mass. 

If the blue giants are binaries it becomes almost impossible to estimate
their mass from their position in the CMD/H-R diagram. Their current position is  
determined by the past mass exchange episodes and various evolutionary
effects. We usually do not have information about the current state of
the companion, nor do we know the initial mass ratio (see also discussion
in Clarke \& Pringle 1992), which would be needed
for the determination of the IMF. Unresolved binaries may easily result in
giving too large a weight to the upper mass bins of the IMF if the star 
consists of two 
(or more) components that initially had roughly the same mass. Unrecognized
neutron-star or white-dwarf companions may lead to an underestimate of the
mass of the visible component. 

If instead the blue giants are evolved stars, the upper mass bins of the IMF
will be underestimated since the masses of the main-sequence progenitors
are likely to have been higher. However, as discussed 
we think we can exclude that these stars are evolved.

If the binary fraction is known, one can make assumptions about the frequency
distribution of different mass ratios and the contribution of these components
to the IMF. Detailed comparisons with the predictions of the models of Pols
\& Marinus (1994) may help to constrain the binary fraction and the types of
mass transfer. However, since stars do not need to be binaries to become Be 
stars and since NGC 330 is so rich in Be stars, we may see the combined
result of two different effects: the rotational distortion and red excess for
single Be stars, and the main-sequence widening and blue (Be) stragglers due to
binarity. At this point, we do not see how these effects can be disentangled. 
Some of the Be stars may indeed be single stars, since Balona (1992)
found a number of $\lambda$ Eri variables among them, which 
are not considered binaries. On the other hand, the majority of
the Galactic Be stars (Slettebak 1988) appear to be Be stars. 

Thus in addition to observational uncertainties and crowding, the 
determination of the IMF of NGC 330 is severely affected by the above 
described effects, probably even more so than other Magellanic Cloud 
clusters because of its high fraction of Be stars.  
As Massey et al.\ (1995) point out, spectral classifications in addition to 
photometry are a prerequisite for any IMF determinations but as we have seen
not even spectroscopy can give all the necessary information.
We are still far from knowing ``typical'' IMF slopes, let alone how the IMF
varies in different (metallicity) environments since especially binary 
fraction and binary mass ratios are unknown.  
Synthetic CMDs, e.g., of Pols \& Marinus may help with these 
problems, however, in the case of NGC 330 
the code would need to be adjusted such that single Be-star evolution and
binary Be-star evolution is taken into account. Even then, reproducing what
we see in NGC 330 would not necessarily lead to conclusive answers since 
there are many free parameters (such as binary fraction,
rotational velocities, initial mass ratios, Be-star fraction and Be lifetimes) 
so that several combinations may reproduce the observed result. Attempts
to reproduce observed CMDs of four young LMC clusters with 
synthetic CMDs assuming binary fractions between 0 and 50\% have been
made by Subramaniam \& Sagar (1995). They note that in some cases, quite
extreme binary fractions are required to reproduce the observations. 
Subramaniam \& Sagar did not account for Be stars in the synthetic CMDs.

\section{The age calibration of Magellanic Cloud clusters}\label{sect_age}

The age calibration of Magellanic Cloud clusters is an important input
parameter for population synthesis models. Ages of Magellanic clusters are
in part based on integrated colours, a rather crude method, and in part
on CMDs obtained photographically or by CCD photometry. Ages obtained from
the CMDs and isochrone fitting are being used to recalibrate ages based on
integrated colours. Integrated colours, in turn, provide an essential input
base for population synthesis models. The properties of young clusters
in particular are important for modeling active galaxies and starburst
galaxies. 

For example,
Bica et al.\ (1990) investigated integrated spectra of LMC and SMC
clusters in the red and far-red (560--1000nm). They dated the
red-supergiant phase to occur between 7 and 12 Myrs. As we
have shown, however, the much older cluster NGC 330 is heavily
dominated (85\% contribution in the $K$ band) by red supergiants
(Grebel 1995, Grebel et al.\ in prep.).  Bica et al.\ (1990) based their age
derivations on the integrated Johnson (\ub)$_0$ and Gunn (u--v)$_0$
colours and the apparent main-sequence turnoff visible in CMDs.
Young Magellanic Clouds clusters with a high Be star content like
NGC 330 may appear bluer in integrated light due to the
UV excess of some of these bright stars,
than one would expect for a cluster
of the same age without Be stars. Furthermore, a contribution to the UV
flux may come from He stars formed in binary interactions (Pols et al.\
1991, Pols 1994, Pols \& Marinus 1994). 
Thus these clusters may be dated younger than they really are. 

The age calibration of young Magellanic Cloud clusters will be off 
if blue stragglers are generally present, as they are in young Galactic 
open clusters. If only photometric data with one colour index are
used, unrecognized blue giants
may place the turnoff region up to a magnitude or more too bright (s.\
Carney et al.\ 1985, Sect.\ \ref{sect_photdata}), 
depending on the age of the cluster. Obviously,
this leads to a systematic underestimation of the cluster age. The exact
age that is derived strongly depends on the parameters of the underlying
stellar models (see Chiosi et al.\ 1995).

A first indicator for blue stragglers is the presence of stars above the
main sequence and below the blue supergiant locus, if isochrones are fitted
such that they include both the main-sequence band and the loci of blue
and red supergiants. In fact, our simultaneous isochrone-fitting technique
Roberts 1996) in several colours does not accept the blue giants 
as part of the main sequence, and this is how we first identified them
as anomalous, without spectroscopy.
In absence of spectroscopy or multi-colour
photometry, blue stragglers will not be recognized for what they are
so that
age spreads will be derived to fit their positions by isochrones as well
(Sect.\ \ref{sect_photdata}). 

The second implication of our findings on the age of NGC 330 is that an
age spread, if present, is small and within the uncertainty of the age 
derived for NGC 330 from isochrone fitting that includes the supergiants.  
The stars in NGC 330 may be considered coeval within $\pm 4$ Myr (Grebel
1995, Grebel et al.\ in prep.).

If blue stragglers are generally present, large age spreads
need no longer to be invoked to explain the morphology of a CMD. 
Lack or presence of a considerable age spread in clusters will have 
important ramifications for theories of star formation.  We believe that
once spectroscopic data are available for a larger set of clusters,
previously inferred age spreads will be much reduced. 

\section{Concluding remarks}\label{sect_conc}

The discovery of blue giants in NGC 330 implies that this cluster
is older than has been derived in several previous studies
(main-sequence turnoff is shifted to
fainter magnitudes) and that the previously inferred large age spread
(e.g., Chiosi et al.\ 1995) is much smaller. The location of the blue
and red supergiants is in agreement with the main-sequence turnoff 
inferred from spectral classifications and isochrone fitting. 

We reject the possibility that these stars are subgiants evolving from the main
sequence towards the red-supergiant phase, since evolution proceeds so
rapidly that it is unlikely to catch stars in this phase. 
Considering that high-resolution spectroscopic studies of blue giants
in NGC 330 did not show enhancement of both He and N and the 
accompanying C and O deficiency, we argue that the blue
giants probably are not evolved stars such as post-red supergiants.  

We argue that 
rotational displacement effects due to rapid rotation and properties 
of Be stars, which is what most of the blue giants are, can account for
their position in the H-R diagram. 
Departures from spherical symmetry as discussed
by Collins (1987) prevent the measurement of global $T_{\rm eff}$ and $\log g$
values and may, depending on the inclination angle, make stars appear as
giants. We find that the simulations of the effects of
binaries on CMDs of young clusters by Pols \& Marinus (1994) describe the
observed CMD and H-R diagram morphology very well. These models account for a 
strongly widened main sequence, the presence of Be stars, and blue unevolved 
stars above the main sequence through binary interactions and mass transfer
that lead to the formation of blue stragglers. 

We suggest that the blue giants are most likely a mixture of (1) rapidly
rotating B/Be stars of varying orientation and (2) blue stragglers 
formed by interactions in binary systems. Both possibilities are 
compatible with these stars being core H burning. 
Faint short-periodic early-type eclipsing variables
and blue straggler binaries have been found in the 
surroundings of NGC 330 (Sect.\ \ref{sect_BS2}), and we suggest additional 
observational tests to quantify the binary presence in NGC 330.  
Isochrones in the observational plane 
considering a certain binary fraction and rotational displacement 
rather than representing low-rotation, single-star
evolution would be most useful. 

We discuss evidence that NGC 330 is not an exception in containing ``blue
giants'', and that blue stragglers are possibly also present in NGC 1818,
NGC 2004, and NGC 2100. 
We argue that unrecognized blue stragglers and binaries can 
severely impair the determination of IMFs. They can also lead to an 
underestimation
of the ages of young Magellanic clusters and to the derivation
of erroneously large age spreads in star formation times, as happened in
the case of NGC 330.  

\section*{Acknowledgements}

EKG gratefully acknowledges valuable discussions with Jon \& Karen Bjorkman,
Steve Cranmer, Katy Garmany, Reinhard Hanuschik, and Rens Waters, who directed 
me to useful references that helped to uncover the probable nature of the
blue giants in NGC 330. We are grateful to Danny Lennon for making available 
his results prior to or during publication.
We thank Klaas de Boer for a critical reading of the
text and our referee, Dr.\ V.\ Caloi, for useful comments. 

EKG was supported through a Graduate Fellowship in the Graduiertenkolleg 
``The Magellanic System, Its Structure, And Its Interaction With The Milky 
Way'' of the German Research Foundation (DFG). WB was supported by the 
German Research Foundation (DFG) under grant Yo 5/16--1.
This research has made use of the Simbad database,
operated at CDS, Strasbourg, France, and of the NASA Astrophysics Data
System, operated at CfA, Harvard, USA.


\begin{thebibliography}{}

\bibitem[]{} 
Arp, H., 1959, AJ 64, 254

\bibitem[]{} Balona, L. 1992, MNRAS 256, 425

\bibitem[]{} 
Barbuy, B., Spite, M., Spite, F., Milone, A. 1991, A\&A 247, 15

\bibitem[]{}
Beech, M., Mihalas, R. 1989, A\&A 213, 127

\bibitem[]{} 
Bessell, M.S. 1991, IAU Symp.\ 148 ``The Magellanic CLouds'',
Eds.\ R.\ Haynes \& D.\ Milne, Kluwer, Dordrecht, p.\ 273

\bibitem[]{} Bessell, M.S., Wood, P.R. 1993,
in ``New Aspects of Magellanic Clouds Research'', Eds.\ B.\ Baschek,
G.\ Klare, J.\ Lequeux, Springer Lecture Notes in Physics 416, p.\ 271

\bibitem[]{}
Bica E., Alloin D., Santos Jr. J.F.C. 1990, A\&A 235, 103

\bibitem[]{}
B\"ohm-Vitense, E., Hodge, P., Proffit, C. 1985, ApJ 292, 130

\bibitem[]{} 
Bomans, D.J., Grebel, E.K. 1994, Space Sci.\ Rev.\ 66, 61

\bibitem[]{}
Brunish, W.M., Truran, J.W. 1982, ApJS 49, 447

\bibitem[]{} 
Caloi, V., Cassatella, A., Castellani, V., Walker, A. 1993,
A\&A 271, 109

\bibitem[]{}
Caloi, V., Cassatella, A. 1995, A\&A 295, 63

\bibitem[]{}
Carney, B.W., Janes, K.A., Flower, P.J. 1985, AJ 90, 1196

\bibitem[]{}
Cassinelli, J.P., Cohen, D.H. 1994, in ``Pulsation, Rotation, and Mass Loss
in Early-Type Stars'', IAU Symp.\ 162, Eds.\ L.A.\ Balona, H. Henrichs, \&
J.M.\ le Contel, Kluwer, Dordrecht, p.\ 189

\bibitem[]{} 
Charbonnel, C., Meynet, G., Maeder, A., Schaller, G., 
Schaerer, D. 1993, A\&AS 101, 415

\bibitem[]{}
Chiosi, C, Summa, C. 1970, Ap.\ Space Sci.\ 8, 478

\bibitem[]{} 
Chiosi, C, Vallenari, A., Bressan, A., Deng, L., Ortolani, S.
1995, A\&A 293, 710 

\bibitem[]{}
Clarke, C.J., Pringle, J.E. 1992, MNRAS 255, 423

\bibitem[]{}
Collier, A.C., Jenkins, C.R. 1984, MNRAS 211, 391

\bibitem[]{}
Collins, G.W. 1987, in ``Physics of Be Stars'', IAU Coll.\ 92,
Eds.\ A.\ Slettebak \& T.P.\ Snow, Cambridge University Press, Cambridge,
p.\ 3

\bibitem[]{}
Collins, G.W., Smith, R.C. 1985, MNRAS 213, 519

\bibitem[]{}
Collins, G.W., Truax, R.J., Cranmer, S.R. 1991, ApJS 77, 541

\bibitem[]{}
Davies, R.D., Elliot, K.H., Meaburn, J. 1979, MemRAS 81, 89

\bibitem[]{}
de Loore, C., Vanbeveren, D. 1994, A\&A 292, 463

\bibitem[]{}
Doom, C. 1985, A\&A 142, 143

\bibitem[]{}
Dufton, P.L., Brown, P.J.F., Fitzsimmons, A., Lennon, D.J. 1990, A\&A 232, 431

\bibitem[]{}
Dufour, R.J. 1984, in ``Structure and Evolution of the Magellanic Clouds'',
IAU Symp.\ 108, Eds.\ S.\ van den Bergh \& K.S.\ de Boer, Dordrecht, Reidel,
p.\ 353

\bibitem[]{}
Eggen, O.J., Iben, I.Jr. 1988, AJ 96, 635

\bibitem[]{} 
Feast, M.W., Black, C. 1980, MNRAS 191, 285

\bibitem[]{}
Fitzpatrick, E.L. 1991, PASP 103, 1123

\bibitem[]{}
Fitzpatrick, E.L., Garmany, C.D. 1990, ApJ 363, 119

\bibitem[]{}
Fitzpatrick, E.L., Bohannan, B. 1993, ApJ 404, 734

\bibitem[]{}
Garmany, C.D., Fitzpatrick, E.L. 1989, in ``Physics of Luminous Blue
Variables'', IAU Coll.\ 113, Eds.\ K.\ Davidson \& H.J.G.L.M.\ Lamers,
Kluwer, Dordrecht, p.\ 83

\bibitem[]{}
Gies, D.R., Lambert, D.L. 1992, ApJ 387, 673

\bibitem[]{}
Grebel, E.K. 1995, PhD Thesis, Bonn University

\bibitem[]{} 
Grebel, E.K., Richtler, T., de Boer, K.S. 1992, A\&A 254, L5

\bibitem[]{}
Grebel, E.K., Roberts, Wm J., van de Rydt, F. 1994a, in ``The Local
Group. Comparative and global properties'', 3rd CTIO/ESO Workshop,
ESO Conference \& Workshop Proceedings No.\ 51, Eds.\ A.\ Layden,
R.C.\ Smith, \& J.\ Storm, La Serena, p.\ 148

\bibitem[]{}  
Grebel, E.K., Roberts, W.J., Will, J.M., de Boer, K.S. 1994b,
Space Sci.\ Rev.\ 66, 65


\bibitem[]{} 
Grebel, E.K., Roberts, W.J., Brandner, W., Moneti, A. 
1996, in prep. 

\bibitem[]{}
Grevesse, N., Anders, E., 1989, in ``Cosmic Abundances of Matter'', AIP
Conf.\ Proc., Ed.\ J.\ Waddington, New York, p.\ 1

\bibitem[]{}
Grossman, S.A., Narayan, R., Arnett, D. 1993, ApJ 407, 284

\bibitem[]{}
Humphreys, R.M., Kudritzki, R.-P., Groth, H.G. 1991, A\&A 245, 593

\bibitem[]{} 
Jacoby, G.H., Hunter, D.A., Christian, C.A. 1984, ApJS 56, 257
(avail.\ by anonymous ftp from pandora.tuc.noao.edu)

\bibitem[]{}
Jasniewicz, G., Th/`evenin, F. 1994, A\&A 282, 717

\bibitem[]{} 
J\"uttner, A. 1993, in ``New Aspects of Magellanic Cloud 
Research'', Eds.\ B.\ Baschek, G.\ Klare, J.\ Lequeux, Lecture
Notes in Physics \# 416, Springer, Heidelberg, p.\ 301

\bibitem[]{}
J\"uttner, A., Stahl, O., Wolf, W., Baschek, B. 1993, in ``New Aspects of
Magellanic Cloud Research'', Eds.\ B.\ Baschek, G.\ Klare, J.\ Lequeux,
Lecture Notes in Physics \# 416, Springer, Heidelberg, p.\ 337

\bibitem[]{}
Kippenhahn, R., Weigert, A. 1967, Z.f.Astrophys.\ 65, 251 

\bibitem[]{}
Kudritzki, R.P., Groth, H.G., Butler, K., Husfeld, D., Becker, S. 1987,
in ESO Workshop on the SN 1987A, European Southern Observatory, Garching,
p.\ 39

\bibitem[]{} 
Landolt, A.U. 1992, AJ 104, 340

\bibitem[]{}
Langer, N., Maeder, A. 1995, A\&A 295, 685

\bibitem[]{}
Lennon, D.J. 1994, Space Sci.\ Rev.\ 66, 127

\bibitem[]{}
Lennon, D.J., Dufton, P.L., Mazzali, P.A., Pasian, F., Marconi, G. 1996,
A\&A, submitted

\bibitem[]{}
Lennon, D.J., Kudritzki, R.-P., Becker, S.T., Butler, K., Eber, F., 
Groth, H.G., Kunze, D. 1991, A\&A 252, 498

\bibitem[]{} 
Lennon, D.J., Mazzali, P.A., Pasian, F., Bonifacio, P., 
Castellani, V. 1994, Space Sci.\ Rev.\ 66, 169

\bibitem[]{}
Lennon, D.J., Mazzali, P.A., Pasian, F., Bonifacio, P.,
Castellani, V. 1995, priv.\ comm.

\bibitem[]{}
Lennon, D.J., Dufton, P.L., Mazzali, P.A., Pasian, F., Marconi, G.
1996, A\&A, submitted

\bibitem[]{}
Luck, R.E., Lambert, D.L., 1992, ApJS 79, 303

\bibitem[]{}
Maeder, A. 1987, A\&A 178, 159

\bibitem[]{}
Maeder, A., Peytremann, E. 1970, A\&A 7, 120

\bibitem[]{}
Maeder, A., Peytremann, E. 1972, A\&A 21, 279

\bibitem[]{}
Massey, P., Lang, C.C., DeGoia-Eastwood, K., Garmany, C.D. 1995, ApJ 438,
188 

\bibitem[]{}
Mazzali, P.A., Lennon, D.J., Pasian, F., Marconi, G., Castellani, V. 1995,
A\&A, submitted

\bibitem[]{}
Meliani, M.T., Barbuy, B., Perrin, M.-N. 1995, A\&A 300, 349

\bibitem[]{}
Mermilliod, J.C. 1982, A\&A 109, 37

\bibitem[]{}
Pols, O.R. 1994, A\&A 290, 119

\bibitem[]{}
Pols, O.R., Cot\'e, J., Waters, L.B.F.M., Heise, J. 1991, A\&A 241, 419

\bibitem[]{}
Pols, O.R., Marinus, M. 1994, A\&A 288, 475

\bibitem[]{} Reitermann, A. 1989, PhD Thesis, Heidelberg

\bibitem[]{} 
Reitermann, A., Baschek, B., Stahl, O., Wolf, B. 1990,
A\&A 234, 109

\bibitem[]{}
Roberts, W.J., 1996, in prep.

\bibitem[]{} 
Roberts, W.J., Grebel, E.K. 1995, A\&AS, 109, 313

\bibitem[]{} 
Robertson, J.W. 1974, A\&AS 15, 261

\bibitem[]{}
Rolleston, W.R.J., Dufton, P.L., Fitzsimmons, A., Howarth, J.D., Irwin, 
M.J. 1993, A\&A 277, 10

\bibitem[]{}
Russell, S.C., Dopita, M.A. 1990, ApJS 74, 93

\bibitem[]{}
Russell, S.C., Dopita, M.A. 1992, ApJ 384, 508

\bibitem[]{} 
Schaller, G., Schaerer, D., Meynet, G., Maeder, A. 1992,
A\&AS 96, 269

\bibitem[]{}
Sebo, K.M., Wood, P.R. 1994, AJ 108, 932


\bibitem[]{}
Slettebak, A. 1988, PASP 100, 770

\bibitem[]{}
Slettebak, A., Kuzma, T.J., Collins, G.W. 1980, ApJ 242, 171

\bibitem[]{} 
Stetson, P.B. 1992, ``User's Manual for DAOPHOT II'', Dominion
Astrophysical Observatory, Victoria

\bibitem[]{} 
Stothers, R.B., Chin, C.-W. 1992, ApJ 390, 136

\bibitem[]{}
Stothers, R.B., Chin, C.-W. 1994, ApJ 421, L91

\bibitem[]{}
Stryker, L. 1993, PASP 105, 1081 

\bibitem[]{}
Subramaniam, A., Sagar, R. 1995, A\&A 297, 695

\bibitem[]{}
Tuchman, Y., Wheeler, J.C. 1990, ApJ 363, 255

\bibitem[]{}
Venn, K.A. 1995, ApJ 449, 839

\bibitem[]{}
Walborn, N.R.1988,in``Atmospheric Diagnostics of Stellar Evolution:
Chemical Peculiarity,Mass Loss,and Explosion'',IAU Coll.108, 
Ed.\ K.Nomoto, Springer, Heidelberg, p.70

\bibitem[]{} 
Walborn, N.R., Fitzpatrick, E.L. 1990, PASP 102, 379
(available by anonymous ftp from astro.princeton.edu)

\bibitem[]{}
Walborn, N.R., Lennon, D.J., Haser, S.M., Kudritzki, R.-P., Voels, S.A.
1995, PASP 107, 104

\bibitem[]{}
Will, J.-M., Bomans, D.J., de Boer, K.S. 1995, A\&A 295, 54


\end{thebibliography}
\end{document}